\documentclass[a4paper,fleqn,usenatbib]{mnras}

\usepackage{savesym}
\usepackage{amsmath}
\savesymbol{iint}
\savesymbol{iiint}
\usepackage{txfonts}
\restoresymbol{TXF}{iint}
\restoresymbol{TXF}{iiint}

\usepackage{graphicx}	% Including figure files
\usepackage{amsmath}	% Advanced maths commands
\usepackage{amssymb}	% Extra maths symbols

\newcommand{\be}{\begin{equation}}
\newcommand{\ee}{\end{equation}}
\newcommand{\bea}{\begin{eqnarray}}
\newcommand{\eea}{\end{eqnarray}}

\title[Cyclic multiverses.]{Cyclic multiverses.}

\author[K. Marosek et al.]{
Konrad Marosek,$^{1,2}$\thanks{e-mail: k.marosek@wmf.univ.szczecin.pl}
Mariusz P. D\c{a}browski,$^{1,3,4}$\thanks{e-mail: mpdabfz@wmf.univ.szczecin.pl} Adam Balcerzak$^{1,4}$\thanks{e-mail: abalcerz@wmf.univ.szczecin.pl}\\
$^{1}$Institute of Physics, University of Szczecin, Wielkopolska 15, 70-451 Szczecin, Poland\\
$^{2}$Chair of Physics, Maritime University, Wa{\l}y Chrobrego 1-2, 70-500 Szczecin, Poland\\
$^{3}$National Centre for Nuclear Research, Andrzeja So{\l}tana 7, 05-400 Otwock, Poland\\
$^{4}$Copernicus Center for Interdisciplinary Studies, S{\l }awkowska 17, 31-016 Krak\'ow, Poland
}

\date{Accepted XXX. Received YYY; in original form ZZZ}

\pubyear{2015}

\begin{document}
\label{firstpage}
\pagerange{\pageref{firstpage}--\pageref{lastpage}}
\maketitle

\begin{abstract}
Using the idea of regularisation of singularities due to the variability of the fundamental constants in cosmology we study the cyclic universe models. We find two models of oscillating and non-singular mass density and pressure (''non-singular'' bounce) regularised by varying gravitational constant $G$ despite the scale factor evolution is oscillating and having sharp turning points (''singular'' bounce). Both violating (big-bang) and non-violating (phantom) null energy condition models appear. Then, we extend this idea onto the multiverse containing cyclic individual universes with either growing or decreasing entropy though leaving the net entropy constant. In order to get an insight into the key idea, we consider the doubleverse with the same geometrical evolution of the two ``parallel'' universes with their physical evolution (physical coupling constants $c(t)$ and $G(t)$) being different. An interesting point is that there is a possibility to exchange the universes at the point of maximum expansion -- the fact which was already noticed in quantum cosmology. Similar scenario is also possible within the framework of Brans-Dicke theory where varying $G(t)$ is replaced by the dynamical Brans-Dicke field $\phi(t)$ though these theories are slightly different.
\end{abstract}


%%%%%%%%%%%%%%%%%%%%%%%%%%%%%%%%%%%%%%%%%%%%%%%%%%

%%%%%%%%%%%%%%%%% BODY OF PAPER %%%%%%%%%%%%%%%%%%

\section{Introduction}

\label{intro}
\setcounter{equation}{0}

One of the most fundamental problems of cosmology is the question about initial conditions which in the standard approach involve the discussion of singularities. The singularities appear as a result of our ignorance about extreme states of the universe, where one faces a blow-up of common physical quantities leading to the termination of the evolution of the universe (at least from our point of view). In the standard case of $\Lambda$CDM cosmology a big-bang initial singularity appears and it marks our lack of knowledge about what was the state of the universe before. In general relativity, a singularity is defined as a region for which one cannot extend a (null or timelike) geodesic. This is known as geodesic incompletness \citep[]{Hawking1973}. However, this definition is very general and some more subtleties on the nature of singularities have been investigated leading to some milder (weak singular) states of the universe than the standard big-bang (strong singular) case
\citep[]{Tipler1977,Krolak1986,Fernandez-Jambrina2004}.

It emerges that there is the whole variety of non-standard (or exotic) singularities  \citep[]{Nojiri2005,limits}. Among them a big-rip (strong) singularity associated with the phantom dark energy \citep[]{Caldwell2002,Dafmmodeboxclsecibrowski2003,Caldwell2003}, further classified as type I, a sudden future singularity (SFS or type II) \citep[]{Barrow1986,Barrow2004,Freese2008,Shtanov2002}, as well as numerous other types such as: generalized sudden future singularities (GSFS), finite scale factor singularities (FSF or type III), big-separation singularities (BS or type IV), $w$-singularities (type V) \citep[]{Dabrowski2009,Fernandez-Jambrina2010}, directional singularities \citep[]{Fernandez-Jambrin2007}, little-rip singularities \citep[]{Frampton2011}, and pseudo-rip singularities \citep[]{Frampton2012}. All these singularities except big-rip and little-rip are weak.

Singularities seem to be an obstacle to construct cyclic models of the universe - something which was a strong ancient Greek culture believe (Phoenix Universe). In relativistic cosmology this idea was first developed by  \citet{Tolman1987} and further studied by numerous authors  \citep[]{Markov1984,Linde1984,Barrow1995,Sahni2015}. However, the crucial issue was that despite one was able to formally construct cyclic models, one could not avoid the problem of initial (strong) singularity. The finite number of cycles just shifts the singularity problem into the past initial cycle. Besides, the physical mechanism of a transition from a cycle to a cycle remained unknown because of singularities joining the cycles.

Recently, this problem has been approached \citep[]{Barrow2004a,Alexander2016} by the application of a possibility that fundamental constants in the universe become dynamical fields \citep[]{Uz2011}. They influence the singularities in a way that these become milder (weaker) or just disappear. So far mainly the theories with dynamical gravitational constant $G$ \citep[]{Brans1961}, the fine structure constant $\alpha$ \citep[]{Barrow1998,Barrow2010}, the electric charge $e$ \citep[]{Bekenstein1982}, and the varying speed of light $c$ \citep[]{Barrow1999,Albrecht1999,Barrow1999a,GOPAKUMAR2001} have been investigated. It has been found that varying $\alpha$ and varying $c$ theories allow the solution of the standard cosmological problems such as the horizon problem, the flatness problem, and the $\Lambda-$problem.

In our previous papers \citep[]{Dabrowski2013,Dabrowski2014} we have applied varying constants theories to regularise (remove or weaken their nature) various types singularities in cosmology. This idea coincides with an idea of regularisation by the higher-order curvature terms \citep[]{Barrow2008,Houndjo2010} which are of quantum (or stringy) origin and obviously they may be equivalent to the effect of the varying constants. In this paper we will further regularise some other cosmological singularities and study the cyclic models of the universe which are due to the variability of the fundamental constants. We will also extend this idea to the multiverse containing individual universes with regularised singularities which have both growing and decreasing entropy though leaving the total entropy constant.

The paper is organised as follows. In Section \ref{models} we present the field equations for varying $c$, $G$, and $\alpha$ universes. In Section \ref{regular} we investigate new example of regularisation of a weak singularity. In Section \ref{cyclic} we discuss regularised by varying $G$ cyclic and non-singular in matter density and pressure models of the universe. In Section \ref{CyclicMulti} we study the multiverse model of constant entropy composed of cyclic individual universes of growing or diminishing entropy. In Section \ref{cyclicBD} we consider similar multiverse model but within the framework of Brans-Dicke theory of varying $G$ replaced by a dynamical field $\phi$. In Section \ref{conclusion} we give our conclusions. 

\section{Varying constants cosmologies}
\setcounter{equation}{0}
\label{models}

We consider Friedmann universes within the framework of varying speed of light $c$ theories and varying gravitational constant $G$ theories. The field equations read as \citep[]{Barrow1999,Albrecht1999,Barrow1999a,GOPAKUMAR2001}
\bea
\label{rho}
\frac{\dot{a}^2}{a^2} &=& \frac{8\pi G(t)}{3} \varrho - \frac{kc^2(t)}{a^2}~,\\
\label{p}
\frac{\ddot{a}}{a} &=& - \frac{4\pi G(t)}{3} \left( \varrho + \frac{3p}{c^2(t)} \right)~,
\eea
where $p$ is the pressure and $\varrho$ is the mass density, and the modified continuity equation is
\be
\label{conser}
\dot{\varrho}(t) + 3 \frac{\dot{a}}{a} \left(\varrho(t) + \frac{p(t)}{c^2(t)} \right) = - \varrho(t) \frac{\dot{G}(t)}{G(t)}
+ 3 \frac{kc(t)\dot{c}(t)}{4\pi Ga^2}~.
\ee
Here $a \equiv a(t)$ is the scale factor, a dot means the derivative with respect to time $t$, $G=G(t)$ is time-varying gravitational constant, $c=c(t)$ is time-varying speed of light, and the curvature index $k=0, \pm 1$. The idea to get Eqs. (\ref{rho}) and (\ref{p}) is that one varies the generalized Einstein-Hilbert action in a preferred frame (since the Lorentz symmetry is violated), and then gets the Einstein field equations with $c$ varying (playing the role of a scalar field) \citep[]{Barrow1999,Albrecht1999,Barrow1999a,GOPAKUMAR2001} (but see also \citep[]{Ellis2005,Ellis2007,Magueijo2008} ). The Eq. (\ref{conser}), on the other hand, can be obtained by direct differentiating (\ref{rho}) and inserting into (\ref{p}).

It is worth mentioning the similarity of varying $c$ equations above with the field equations for Friedmann universes with varying fine structure constant $\alpha$ (or alternatively, the electron charge $e$ \citep[]{Bekenstein1982}) which are as follows \citep[]{Barrow2004a,Alexander2016}
\bea
\label{alrho}
\frac{\dot{a}^2}{a^2} & = & \frac{8\pi G}{3} \left( \varrho_r + \varrho_{\psi} \right) - \frac{kc^2}{a^2},\\
\label{alp}
\frac{\ddot{a}}{a} & = & -\frac{8\pi G}{3} \left( \varrho_r + 2\varrho_{\psi} \right),
\eea
and
\be
\label{alpsi}
\ddot{\psi} + 3 \frac{\dot{a}}{a} \dot{\psi} = 0,
\ee
where $\varrho_r \propto a^{-4}$ stands for the density of radiation and
\be
\label{alEOS}
\varrho_{\psi} = \frac{p_{\psi}}{c^2} = \frac{\epsilon}{2} \dot{\psi}^2
\ee
stands for the density of the scalar field $\psi$ (standard with $\epsilon = +1$ and phantom with $\epsilon$ = -1 \citep[]{Caldwell2002}) responsible for the variation of the fine structure constant
\be
\label{alpha}
\alpha(t) = \alpha_0 e^{2 \psi(t)} .
\ee
It is easy to notice that (\ref{alpsi}) integrates to $\dot{\psi} = C/a^3$.
Also, one should mention that bearing in mind (\ref{alp}) and (\ref{alEOS}) one cannot get the universe with a bounce in the standard sense (a ''non-singular bounce'') unless $\epsilon = -1$. In other words $\ddot{a} < 0$ throughout the time of the evolution. Putting this other way, for $\epsilon = +1$ there is no way to go from a contraction where $H = \dot{a}/a < 0$ to an expansion where
$H=\dot{a}/a >0$ through a singularity since H should have a minimum while in general one has
\be
\label{Hminimum}
\dot{H} = \left( \frac{\dot{a}}{a} \right)^{\cdot} = - 4\pi G \left( \varrho + \frac{p}{c^2} \right) < 0,
\ee
if the null energy condition is fulfilled. However, a ''singular bounce'' where there is no change of the sign of the second derivative of the scale factor is still possible, and will be discusses in the next sections.

Another point is that the theory represented by the Eqs. (\ref{rho}), (\ref{p}), and (\ref{conser}) is not the same as the Brans-Dicke theory in which $G$ is replaced by a dynamical scalar field $\phi(t)$ which possesses an extra field dynamics equation of motion (analogue of (\ref{alpsi}) in varying $\alpha$ theory) and such a system is stronger constrained than (\ref{rho})-(\ref{conser}) (see our Section \ref{cyclicBD}). One of the points is that in both Brans-Dicke 
\citep[]{sakai2001,Bhattacharya2015} and $G(t)$ theories \citep[]{Mersini2014,Sidharth2015} based on  
(\ref{rho})-(\ref{conser}) black holes other than those of general relativity cannot exist (''no-hair Brans-Dicke theorem") though it is not necessarily the case in varying speed of light theories \citep[]{magueijo2001}. 

\section{Regularising weak singularities and facing a little-rip}
\setcounter{equation}{0}
\label{regular}

In \citet[]{Dabrowski2013} we have shown how regularisation of the series of standard and exotic singularities \citep[e.g.][]{Dabrowski2010} by varying constants was possible. In particular, it applied to sudden future singularities which due to regularisation could be gone through in a similar fashion as in \citet{Dabrowski2011}. In fact, their weak nature was considered by \citet{Fernandez-Jambrina2004,Fernandez-Jambrina2006,Fernandez-Jambrin2007,Fernandez-Jambrina2010}.

Here we use the scale factor which after appropriate choice of parameters admits big-bang, big-rip, sudden future, finite scale factor and $w$-singularities and reads as \citep[]{Dabrowski2013}
\be \label{newscalef} a(t) = a_s \left( \frac{t}{t_s} \right)^m \exp{\left( 1 - \frac{t}{t_s} \right)^n} \equiv a_{BB} a_{ex}~, \ee
with the constants $t_s, a_s, m, n$ \citep[e.g.][]{Barrow1986,Barrow2004,Freese2008,Shtanov2002}, where we have split the scale factor into the two factors, one giving a big-bang singularity $a_{BB}$, and another giving an exotic singularity $a_{ex}$. We modify (\ref{newscalef}) into
\be
\label{amod1}
a_{ex}=a_{s} \exp {\left| 1  - \frac{t}{t_{s}} \right|}^n ~~,
\ee
which allows to extend the scale factor onto the times larger than $t_s$ ($t>t_s$) \citep[]{Barrow2013}:
\be
a_{ex2}(t)=a_{s} \exp {\left(\frac{t}{t_{s}} -1 \right)}^n ~~.
\ee
From (\ref{rho})-(\ref{p}) and (\ref{amod1}) we have
\bea
\label{rhoex1}
{\rho}_{ex1}&=&\frac{3}{8 \pi G \left( t \right)} {\left(1-\frac{t}{t_{s}} \right)}^{2 \left( n -1 \right)} \\
\label{pex1}
p_{ex1} \left( t \right) &=&-\frac{c^2}{8 \pi G \left( t \right)} \left[ 3 \frac{n^2}{{t_{s}}^2}  {\left(1-\frac{t}{t_{s}} \right)}^{2 \left( n -1 \right)} \right. \nonumber \\
&+& \left. 2 \frac{n \left( n-1 \right)}{{t_{s}}^2} {\left(1-\frac{t}{t_{s}} \right)}^{n-2} \right],
\eea
for $0<t<t_s$, and
\bea
{\rho}_{ex2}&=&\frac{3}{8 \pi G \left( t \right)} {\left(\frac{t}{t_{s}}-1 \right)}^{2 \left( n -1 \right)} \\
p_{ex2} \left( t \right) &=&-\frac{c^2}{8 \pi G \left( t \right)} \left[ 3 \frac{n^2}{{t_{s}}^2}  {\left(\frac{t}{t_{s}}-1 \right)}^{2 \left( n -1 \right)} \right. \nonumber \\
&+& \left. 2 \frac{n \left( n-1 \right)}{{t_{s}}^2} {\left(\frac{t}{t_{s}}-1 \right)}^{n-2} \right],
\eea
for $t>t_s$.
There are various cases of weak singularities: for $1 \leq n \leq 2$ there is a sudden future singularity, for $0 \leq n \leq 1$  there is a finite scale factor singularity, and for $ n \geq 2$ there is a generalised sudden future singularity.

Following \citet{Dabrowski2013} we may assume the variability of $G$ as
\be
G \left( t \right) = G_{0} \left| 1 - \frac{t}{t_{s}} \right|^{-r} ,
\label{Gmid}
\ee
which allows the energy density (\ref{rhoex1}), and the pressure (\ref{pex1}) be regularized for $r>2-n$ at $t=t_s$, and the universe may continue its evolution until $t \rightarrow \infty$, where the scale factor diverges $(a \rightarrow \infty)$ and both the mass density $ \rho \rightarrow \infty $ and pressure $ \left| p \right| \rightarrow \infty $ diverge, achieving a (strong) little-rip singularity \citep[]{Frampton2011} (which is a version of a (strong) big-rip \citep[]{Caldwell2003} being reached in an infinite time). The regularisation of a sudden singularity due to (\ref{Gmid}) happens at the expense of getting an infinitely strong gravitational coupling ($G \to \infty$) at $t=t_s$, which is the characteristic being explored later in the paper. 

Another option to regularise a sudden future singularity (which admits a blow-up of pressure in (\ref{pex1}) is to use the variability of the speed of light
\be
 c= c_{0} {\left|1- \frac{t}{t_{s}} \right|}^{\frac{p}{2}}
\label{cmid}
\ee
which happens for $ p >2-n $ at $t=t_s$ though at the expense of light velocity slowing down to zero ($ c \rightarrow 0 $) at $t=t_s$. As in the previous case, after going through a regularised singularity, one faces a little rip for $ t \rightarrow \infty $, where $ a \rightarrow \infty $, $ \rho \rightarrow \infty $, and $ \left| p \right| \rightarrow \infty $. Unfortunately, the little rip is a strong singularity and so we cannot continue the evolution of the universe further in order to get the cyclic evolution as in  \citep[]{Tolman1987,Barrow1995}. However, there is a transition through a sudden singularity as being a weak singularity which is regularised by varying constants (\ref{Gmid}) and (\ref{cmid}).

In summary, we have considered the scenario in which the universe starts with a constant scale factor with $m=0$ in (\ref{newscalef}) (a bounce), goes through a sudden singularity, and then reaches another singularity -- a little rip -- found by \citet[]{Frampton2011}. However, in this case we cannot continue the evolution towards the cyclic evolution since the little rip is a strong singularity. We will then try to find solutions which can bring the viable cyclic scenario.

\section{Regularising strong singularities by varying $G$}
\setcounter{equation}{0}
\label{cyclic}

As one can see from (\ref{rho})-(\ref{p}) and (\ref{alrho})-(\ref{alp}), the variability of the fundamental constants admits more freedom to a system of cosmological equations. This eases a construction of the regularised cyclic models of the universe which will be shown below.

\subsection{sinusoidal pulse model - regularising big-bang}

\begin{figure}
%[htbp]
\includegraphics[width=8.3cm]{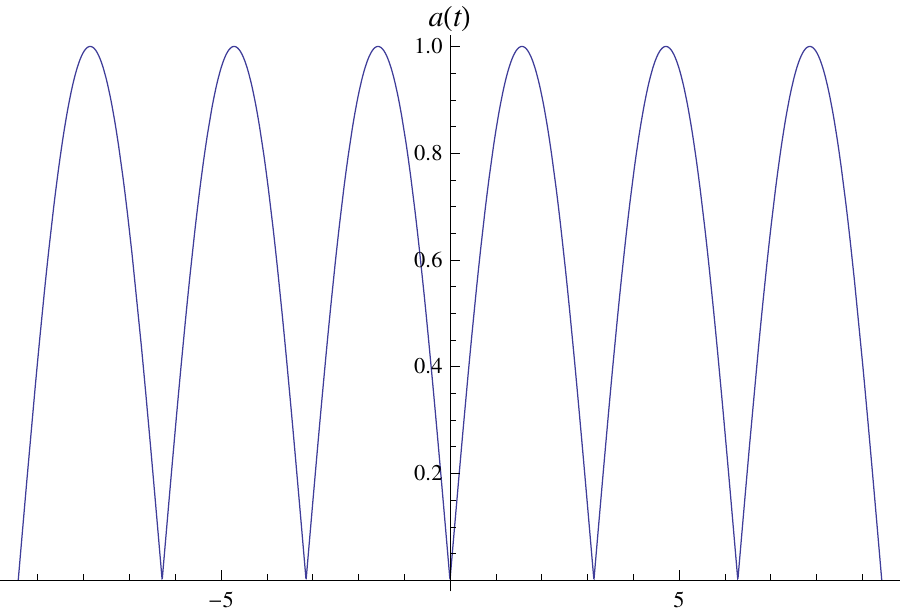}
\includegraphics[width=8.3cm]{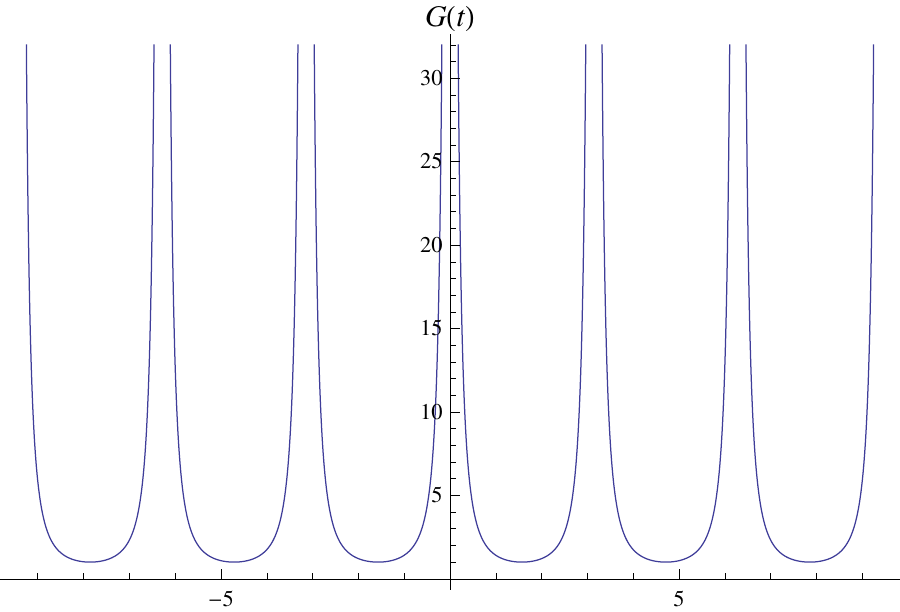}
\caption{The cyclic scale factor (\ref{acyclic1}) and the cyclic gravitational constant (\ref{Gcyclic1}).}
\label{scaleFsin1}
\end{figure}

In this section we use the equations (\ref{rho}) and (\ref{p}) for positive curvature $(k=+1)$ and $\dot{c}=0$ in order to construct a fully regularised (in mass density and pressure) cyclic universe assuming that the scale factor takes the form
\be
\label{acyclic1}
a(t)= a_{0} \left| \sin \left( \pi \frac{t}{t_{c}} \right) \right|
\ee
with $a_0=$const., so the scale factor is ''singular'' (reaching zero as in the big-bang scenario). The main issue here is an additional assumption that the gravitational constant varies as (the speed of light $c$ is taken to be constant here)
\be
\label{Gcyclic1}
G \left( t \right)= \frac{G_{0}}{a^2(t)}.
\ee
The plots of (\ref{acyclic1}) and (\ref{Gcyclic1}) are given in Fig. \ref{scaleFsin1}. It is easy to notice that the standard turning points with curvature singularities (zeros of the scale factor) are regularized by the infinitely strong gravitational coupling ($G \to \infty$).

\begin{figure}
%[htbp]
\includegraphics[width=8.3cm]{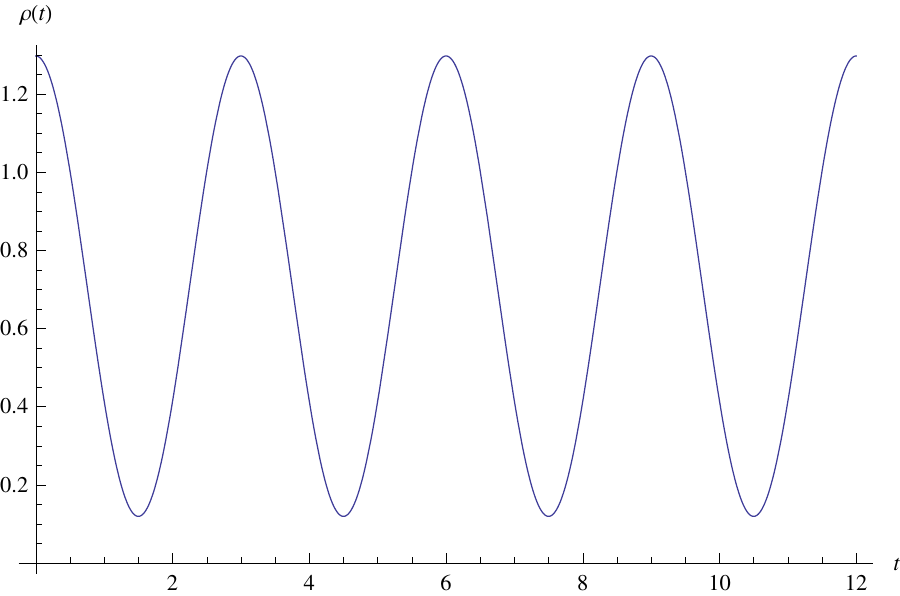}
\includegraphics[width=8.3cm]{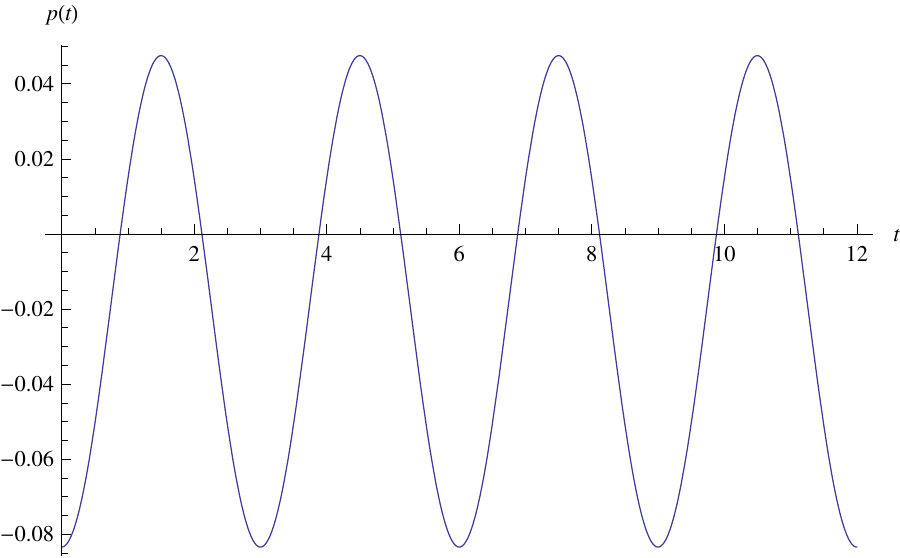}
\caption{The nonsingular mass density (\ref{cyclicr}) and pressure (\ref{cyclicp}) for the models (\ref{acyclic1}) and (\ref{Gcyclic1}).}
\label{rhopsin1}
\end{figure}

This can be seen by inserting (\ref{acyclic1}) and (\ref{Gcyclic1}) into (\ref{rho}) and (\ref{p}) after which we obtain that the mass density $\rho(t)$ and the pressure $p(t)$ are nonsingular and oscillatory (see Fig. \ref{rhopsin1})
\bea
\label{cyclicr}
\rho \left( t \right) &=&  \frac{3}{8 \pi G_{0}} \left[ \frac{\pi ^2 a_{0}^2 \cos ^2 \left( \pi \frac{t}{t_{c}} \right)}{t_{c}^2} +c^2 \right] \geq 0, \\
\label{cyclicp}
p \left( t \right) &=& -\frac{c^2}{8 \pi G_{0}} \left[ \frac{3 \pi ^2  a_{0}^2 \cos ^2 \left( \pi \frac{t}{t_{c}} \right)}{t_{c}^2} +c^2  - 2  \frac{\pi ^2  a_{0}^2}{t_{c}^2} \right] ~~.
\eea
Similar nonsingular in mass density and pressure cosmologies, though also with oscillating non-zero scale factor (a ''non-singular bounce''), driven by the balance between the domain walls and the negative cosmological constant were obtained in \citet[]{Dabrowski1996} and recently reconsidered as the ''simple harmonic oscillator" (SHO) universe \citep[]{Mithani2012,Mithani2014,Graham2014}.
Introducing
\be
p_a = \frac{ c^2}{4 \pi G_{0}} \left( \frac{\pi ^2  a_{0}^2 }{t_{c}^2} + c^2 \right) = p(m t_c) \geq 0,
\ee
where ($m = 0, 1, 2, 3 ...$), one gets the equation of state for (\ref{cyclicr}) and (\ref{cyclicp}) in the form
\be
\label{nullspulses}
p(t) + \rho(t) c^2 = p_a \geq 0,
\ee
which tells us that the null energy condition is always fulfilled (a ''singular bounce'' in the sense of (\ref{Hminimum})). The same is true for the weak energy condition which additionally requires (\ref{cyclicr}) ($\rho \geq 0$). For the strong energy condition we have
\bea
\rho \left( t \right)c^2 +3 p \left( t \right) &=& \frac{3 {a_0}^2 \pi c^2}{4 G_{0} {t_c}^2} \sin^2 \left( \pi \frac{t}{t_{c}} \right) \ge 0 ,
\eea
plus  the condition (\ref{cyclicr}) which means that this condition is fulfilled. In view of (\ref{p}) it shows that the universe is decelerating $(\ddot{a} < 0)$. In general, the strong energy condition means that gravity is attractive, but in our case its atractivity is overbalanced by the strong coupling of $G(t)$ at the singularity ($G \to \infty$).
Since
\be
\rho \left( m t_c \right) = \frac{3 }{8 \pi G_{0}} \left( \frac{\pi ^2  a_{0}^2 }{t_{c}^2} + c^2 \right) = {\rm const.},
\ee
then it is easy to notice that at $t = 0$ the equation of state is of the network of cosmic strings type
\be
\label{EOSbeg}
p \left( 0 \right) = - \frac{1}{3} \rho \left( 0 \right)c^2.
\ee
At the turning points, we have:
\bea
\rho \left( n t_c \right) &=& \frac{3 c^2 }{8 \pi G_{0}} = {\rm const.} ,\\
p\left(n t_c \right) &=& - \frac{c^2}{8 \pi G_{0}} \left( c^2 - \frac{2\pi^2 a_0^2}{t_c^2} \right),
\eea
where $n = 1/2, 3/2, 5/2 \ldots$, and the equation of state there is given by
\be
\label{EOSturn}
p\left(n t_c\right) = - \frac{1}{3} \rho \left( n t_c \right) \left( c^2 - \frac{2\pi^2 a_0^2}{t_c^2} \right).
\ee
In other words, the equation of state varies from the form of the cosmic strings $(w = - 1/3)$ given by (\ref{EOSbeg}) to the form (\ref{EOSturn}) which depends on the parameters of the model $a_0$, $t_c$, and $c$ (cf. Fig. \ref{rhopsin1}). The scale factor is zero for $t = m t_c$ and the big-bang singularity in each cycle is regularized ($a \to 0$, $\varrho \to$ const., $p \to$ const.) at the expense of having strong gravitational coupling $G \to \infty$ at $t=m t_c$.

Bearing in mind the form of the scale factor (\ref{acyclic1}), we have
\bea
\label{dota}
\dot{a}(t) &=& \left\{
\begin{array}{l}
\frac{\pi a_0}{t_c} \cos{\pi \frac{t}{t_c}},\hspace{0.5cm}mt_c \leq t \leq (m+1) t_c, \\
\\
-\frac{\pi a_0}{t_c} \cos{\pi \frac{t}{t_c}},\hspace{0.5cm}(m+1)t_c \leq t \leq (m+2) t_c, \nonumber \\
\end{array}
\right. \
\eea
which means that it both may increase $(\dot{a} > 0)$ and decrease $(\dot{a} < 0)$. Same refers to  the Hubble parameter
\be
H = \frac{\dot{a}}{a} = \frac{\pi}{t_c} \cot{\pi \frac{t}{t_c}},
\ee
which is positive for $mt_c \leq t \leq (m+1) t_c$ and negative for $(m+1)t_c \leq t \leq (m+2) t_c$. The second derivative of the scale factor fulfills a simple harmonic oscillator equation
\be
\ddot{a} = - \frac{\pi^2}{t_c^2} a(t)
\ee
so that it is always negative (deceleration) and this is exactly what we see in the upper panel of Fig. \ref{scaleFsin1}. No turning point of evolution for the scale factor $a(t)$ is possible (a ''singular bounce'' in the sense of the scale factor, but a ''non-singular'' bounce in the sense of the matter density $\varrho$ and pressure $p$).

In fact, from (\ref{nullspulses}) we can conclude that the null energy condition is fulfilled for the models under study. However,  from the second Einstein equation (acceleration equation) we can easily see that
\be
\label{dotH}
\dot{H} - \frac{\dot{a}}{a} \frac{c^2}{a^2} = \dot{H}_{eff} = - 4 \pi G \left( \rho + \frac{p}{c^2} \right) \leq 0 ,
\ee
and so there is no possibility for the universe to ''bounce'' i.e. to change from $H = \dot{a}/a < 0$ (contraction) to
$H = \dot{a}/a > 0$ (expansion) since $H$ should have a minimum. In our model $G$ is varying and this is why we have an extra term on the left-hand side of (\ref{dotH}) , but it does not help a proper "bounce" for the scale factor (geometry) since
\be
\dot{H} = - \frac{\pi^2}{t_c^2} - \frac{\dot{a}^2}{a^2}  < 0 .
\ee
Defining
\be
H_G = \frac{\dot{G}}{G} = - 2 \frac{\dot{a}}{a} = - 2 H
\ee
we can see that $\dot{H}_G > 0$ always accompanies $\dot{H} < 0$ (a contraction of space is always balanced by an expansion of gravity) which makes an apparently singular and sharp bounce regular in matter density and pressure due to the special form of running gravity.

Similar situation appears in ekpyrotic/cyclic scenarios rooted in superstring/brane theories \citep[]{Khoury2001,Khoury2002,Steinhardt2002,Khoury2004,Donagi2001,Turok2005}, where one has some special coupling of gravity in the Lagrangian of a 4-dimensional theory (Einstein frame) as below
\be
S = \int d^4 x \sqrt{-g} \left[ \frac{c^4}{16\pi G} R   - \frac{1}{2}
\partial_{\mu}\phi \partial^{\nu} \phi - V(\phi) +
\beta^4(\phi) (\varrho_R + \varrho_m) \right]~,
\ee
where the potential has an explicit form as
\be
V(\phi) = V_0 \left(1 - e^{-s\phi} \right) F(\phi), \hspace{0.5cm}
F(\phi) \propto e^{-\frac{1}{g_s}}~,
\ee
and $\varrho_R$ is the energy density of radiation while $\varrho_m$
is the energy density of matter, $g_s$ - dilaton/string coupling constant, $\phi$ the scalar field, $\beta(\phi)$ the scalar field coupling, and $s, V_0$ are constants.

This scenario is admissible only for the collision of the boundary
branes and composes of infinitely many such collisions - each of
them produces a big-bang after which the evolution repeats in contrast to a one
brane collision in the earlier ekpyrotic scenario.

The idea of a cyclic universe there is that the 5th dimension
(orbifold) collapses, while the 4-dimensional theory has no
singularity. This can be seen effectively by looking onto the
generalised Friedmann equations which read as
\bea
H^2 &=& \frac{8\pi G}{3} \left(\frac{1}{2} \dot{\phi}^2 + V +
\beta^4 \varrho_R + \beta^4 \varrho_m \right)~,\\
\frac{\ddot{a}}{a} &=& -\frac{8\pi G}{3} \left(\dot{\phi}^2 - V +
\beta^4 \varrho_R + \frac{1}{2} \beta^4 \varrho_m \right)~.
\eea
The crucial point is that one chooses appropriate form of coupling $\beta \propto 1/a$, and we have
\bea
\label{rorb}
\varrho_R \sim \frac{1}{[a\beta(\phi)]^4} \sim \frac{1}{\left(a
\frac{1}{a} \right)^4} \sim {\rm const.}~,\\
\label{romb}
\varrho_m \sim \frac{1}{[a\beta(\phi)]^3} \sim \frac{1}{\left(a
\frac{1}{a} \right)^3} \sim {\rm const.}~,
\eea
in the limit $a \to 0$, where the standard big-bang singularity
usually appears.

In our approach, the role of the coupling $\beta(\phi) \propto 1/a$ is played by the running gravitational constant $G(t) \propto 1/a^2(t)$  which regularises the mass density and pressure analogously as in (\ref{rorb}) and (\ref{romb}) of cyclic scenarios. So one has analogously a kind of ''singular bounce'' in the scale factor $a(t)$ and a ''non-singular'' bounce in the mass density and pressure.

\subsection{tangential pulse model - regularising big-bang and big-rip}

\begin{figure}
%[htbp]
\includegraphics[width=8.3cm]{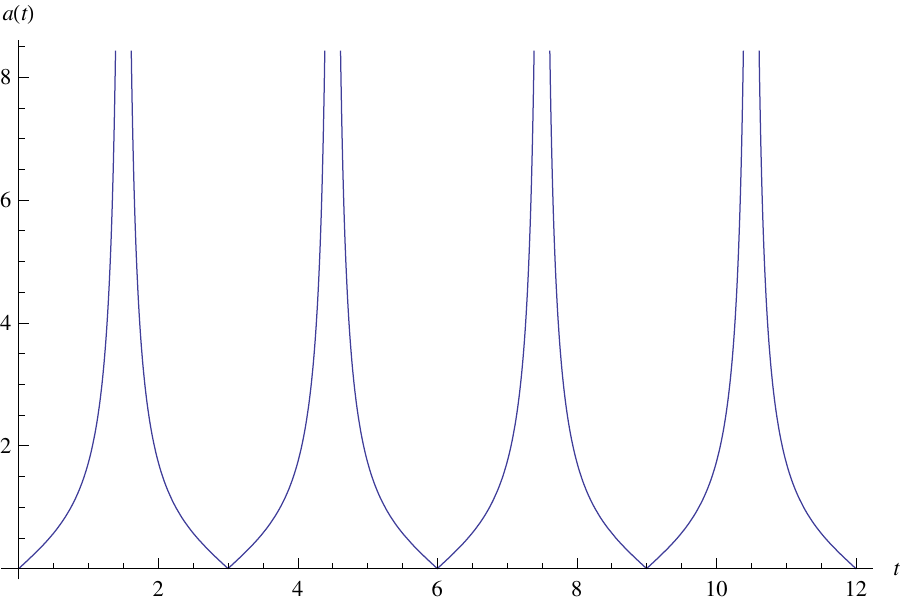}
\includegraphics[width=8.3cm]{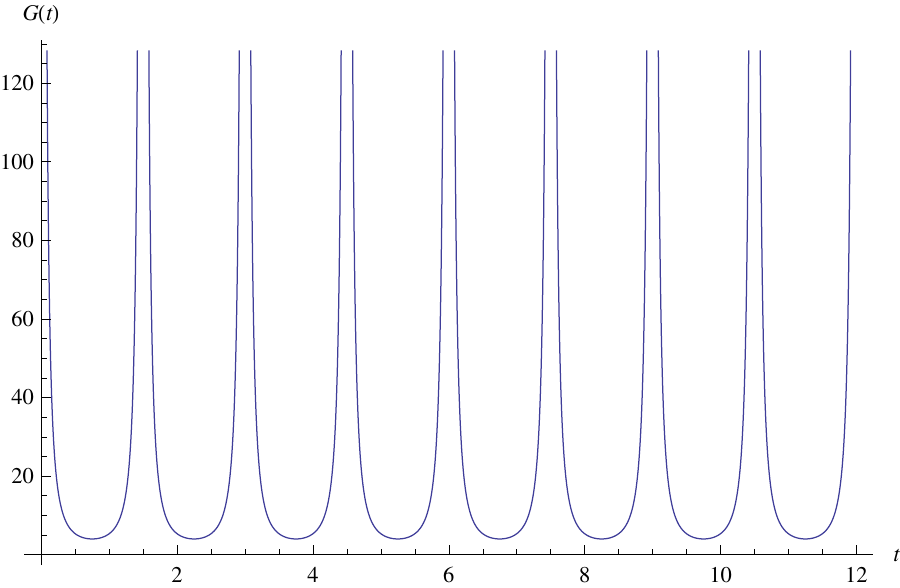}
\caption{The cyclic scale factor (\ref{acyclic2}) and gravitational constant (\ref{Gcyclic2}).}
\label{tanag}
\end{figure}

Another interesting case is possible when one chooses the scale factor to be
\be
\label{acyclic2}
a(t)= a_{0} \left| \tan \left( \pi \frac{t}{t_{s}} \right) \right|~~,
\ee
and the gravitational constant to vary as
\be
\label{Gcyclic2}
G \left( t \right) = \frac{4G_s}{\sin^2{\left( 2 \pi \frac{t}{t_s} \right)}} ~~.
\ee
We can see that the scale factor (\ref{acyclic2}) is infinite ($a \to \infty$) for $t = n t_s$ with $n=1/2,3/2,5/2, \ldots$ (like at big-rips) and it is zero for $t=mt_s$ with $m=0,1,2,3, \ldots$ (like at big-bangs) so that we can say that we face ''singular bounces'' (see Fig.\ref{tanag}). Besides, each time the scale factor $a(t)$ attains a singular value (vanishes or reaches infinity), the gravitational coupling becomes infinite ($G \to \infty$). This is somewhat strange since it regularises both big-bang and big-rip so its ''attractive'' or ''repulsive'' nature does not seem to have any difference here. 

\begin{figure}
%[htbp]
\includegraphics[width=8.3cm]{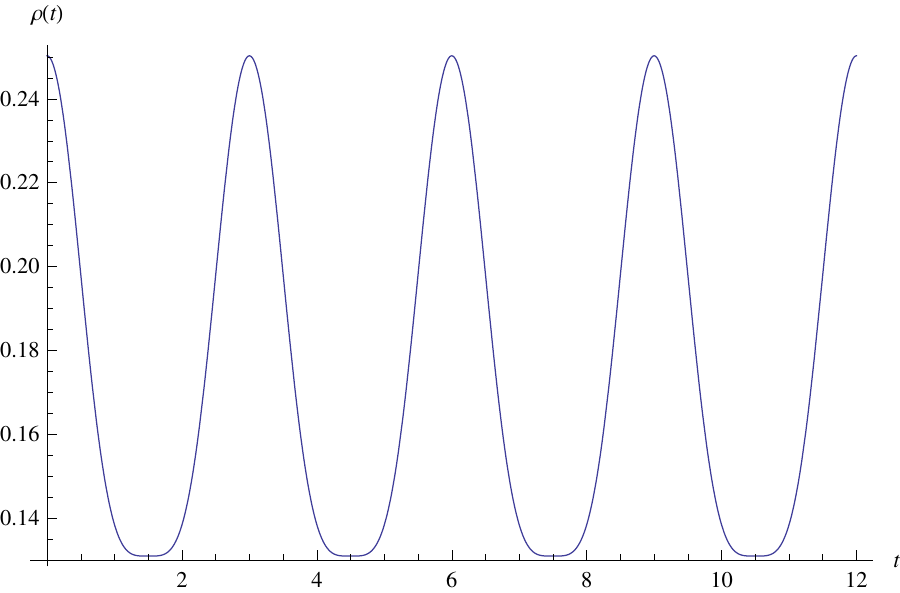}
\includegraphics[width=8.3cm]{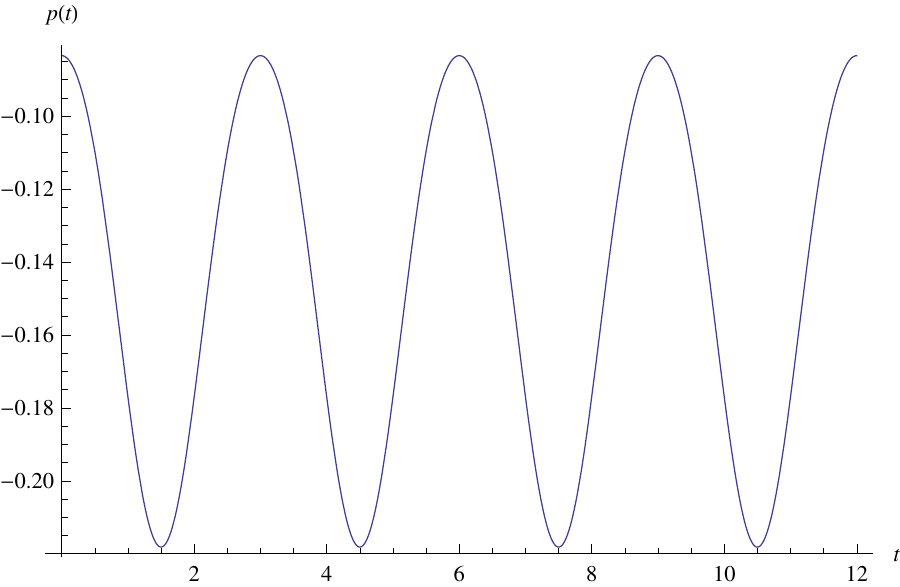}
\caption{The nonsingular mass density (\ref{rcyclic2}) and pressure (\ref{pcyclic2}) for the models (\ref{acyclic2}) and (\ref{Gcyclic2}).}
\label{rhopsin2}
\end{figure}

\begin{figure}
%[htbp]
\includegraphics[width=8.3cm]{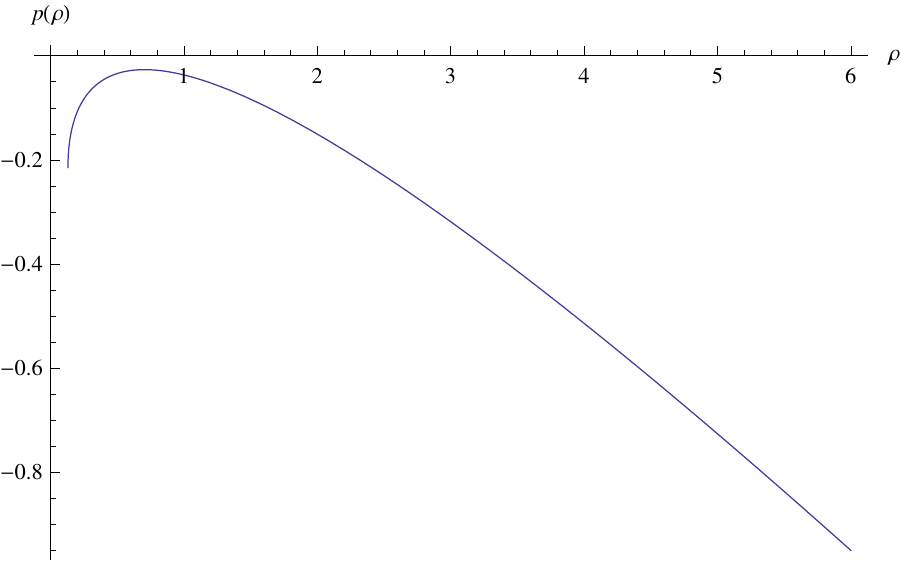}
\caption{The equation of state (\ref{prho}) for the model (\ref{acyclic2})-(\ref{Gcyclic2}). The pressure is always negative and reaches the same initial value (\ref{rhomin}) for $t = nt_s$ and $n = 1/2, 3/2,5/2,...$.}
\label{plotprho.pdf}
\end{figure}

The mass density and pressure are given by
\bea
\label{rcyclic2}
\rho \left( t \right) &=&  \frac{3}{8 \pi G_{s}} \left[ \frac{\pi ^2}{t_{s}^2}+ \frac{3 c^2  \cos ^4 \left( \pi \frac{t}{t_{s}} \right)}{a_0^2} \right]  \geq 0,  \\
\label{pcyclic2}
p \left( t \right) &=& -\frac{c^2}{8 \pi G_{s}} \left[  \frac{\pi ^2}{t_{s}^2}+4 \frac{ \pi^2 \sin ^2 \left( \pi \frac{t}{t_{s}} \right)}{t_{s}^2} +  \frac{ c^2  \cos ^4 \left( \pi \frac{t}{t_{s}} \right)}{a_0^2}  \right], \nonumber \\
\eea
from which it is clear that they are regular (cf. Fig.\ref{rhopsin2}) - in that sense, we have regularised big-rips and big-bangs (''non-singular bounces''). From (\ref{rcyclic2}) and (\ref{pcyclic2}), one calculates that the equation of state is
\be
\label{prho}
p  \left( \rho  \right) = - c^2 \left[ \frac{\pi}{2 G_{s} {t_s}^2}+\frac{\rho}{3}-\frac{{\pi}^{\frac{3}{2}} a_0}{c {t_s}^2} \sqrt{\frac{2 \rho}{3 G_{s}} - \frac{\pi}{4 {G_{s}}^2 {t_s}^2}}  \right],
\ee
where
\be
\label{rhomin}
\rho \geq \rho_{min} = \varrho(nt_s) = \frac{3\pi}{8G_s t_s^2} ,
\ee
which is in agreement with (\ref{rcyclic2}) for $t = n t_s$. The maximal values of the energy density and pressure are given by
\bea
\varrho_{max} = \varrho(mt_s) = \frac{3}{8\pi G_s} \left( \frac{\pi ^2}{t_{s}^2}+ \frac{3 c^2}{a_0^2} \right) ,\\
p_{max} = p(mt_s) = -\frac{c^2}{8 \pi G_{s}} \left(  \frac{\pi ^2}{t_{s}^2}+  \frac{ c^2}{a_0^2} \right) .
\eea
The plot of (\ref{prho}) is given in Fig.\ref{plotprho.pdf}.

Bearing in mind (\ref{rcyclic2}) and (\ref{pcyclic2}), one sees that the null energy condition
\be
c^2 \rho \left( t \right)+p \left( t \right) = \frac{c^2}{4 \pi G_{s}} \left[ \frac{\pi ^2}{t_{s}^2} - 2 \frac{ \pi^2 \sin ^2 \left( \pi \frac{t}{t_{s}} \right)}{t_{s}^2} +  4 \frac{ c^2  \cos ^4 \left( \pi \frac{t}{t_{s}} \right)}{a_0^2}  \right]
\ee
is satisfied at $t = mt_s$ and violated at $t=nt_s$. We can then say that in this context (cf. the discussion around the formula 
(\ref{Hminimum})) we have ''singular bounces'' at big-bangs and ''non-singular bounces'' at big-rips (null energy condition is violated by phantom). It proves that the former is a big-rip-like singularity and the latter is a big-bang-like singularity. The word "like" comes from the fact that the mass density and pressure are regular which is not the case at a big-bang and a big-rip. Perhaps better names would be "regularised" big-rip and "regularised" big-bang. It is interesting to note that the pressure at both of these situations $p(mt_s)$ and $p(nt_s)$ is negative.

The strong energy condition reads as
\be
c^2 \rho \left( t \right)+3 p \left( t \right) =  \frac{3 c^2}{4 \pi G_{s}}  \left[ \frac{ c^2  \cos ^4 \left( \pi \frac{t}{t_{s}} \right)}{a_0^2} -2 \frac{ \pi^2 \sin ^2 \left( \pi \frac{t}{t_{s}} \right)}{t_{s}^2}   \right]
\ee
so that
\bea
c^2 \rho \left( m t_s \right) + 3 p \left( m t_s \right) &=& \frac{3 c^4}{4 \pi G_{s} {a_0}^2} \ge 0 , \\
c^2 \rho \left( n t_s \right) + 3 p \left( n t_s \right) &=& - \frac{3 c^2}{2 \pi G_{s} {t_s}^2} \leq 0 .
\eea
Further, bearing in mind that
\be
p(nt_s) = - \frac{5\pi c^2}{8 G_s t_s^2} , \hspace{0.5cm} c^2 \varrho(nt_s) =  \frac{3 \pi c^2}{8 G_s t_s^2},
\ee
one concludes that the equation of state is that of phantom
\be
\label{phaneos}
p(nt_s) = - \frac{5}{3} \varrho(nt_s) c^2 .
\ee
An interesting fact for the model with such an equation of state (though for the standard Einstein relativity) was found by \citet[]{Fernandez-Jambrina2004}. In fact, it does not lead to the null geodesics incompleteness. In our case one has it effectively due to the influence of the special gravitational coupling of $G(t)$ given by (\ref{Gcyclic2}). It is also interesting to note that the model of similar equation of state violates the conditions of the Green and Wald theorem of averaging local inhomogeneities in the universe which predict the gravitational radiation contribution only \citep{Green2011,Green2013}.

%\bibitem{GreenWald2011} S.R. Green and R.M. Wald, Phys. Rev. D{\bf 83}, 084020 (2011).

%\bibitem{GreenWald2013} S.R. Green and R.M. Wald, Phys. Rev. D{\bf 87}, 124037 (2013).

Finally, from (\ref{phaneos}) one sees that the dominant energy condition is violated as well for $t=nt_s$. This proves that one deals with phantom-like behaviour of the model at $t=nt_s$.

In this tangential pulse model again no turning point of the evolution for the scale factor $a(t)$ is possible (a ''singular bounce'' in the sense of the scale factor, but a ''non-singular'' bounce in the sense of the matter density $\varrho$ and pressure $p$).

\section{Cyclic multiverse}
\setcounter{equation}{0}
\label{CyclicMulti}

In this section we start with the first law of thermodynamics (conservation of energy) allowing the fundamental constants to vary and extend our discussion of regularised universes onto the model of the multiverse as a set of complementary universes evolving in a way which fulfils the condition of keeping the total entropy constant or growing i.e. obeying the second law of thermodynamics (entropy law).

\subsection{thermodynamics of varying $c$}

We first consider the varying $c$ models ($\dot{G}=0$) starting from the first law of thermodynamics
\be
\label{1stlaw}
dE=TdS-pdV,
\ee
where $E$ is the internal energy, $T$ the temperature, $S$ the entropy, $p$ the pressure, and $V$ the volume.
Since
\be
\label{mc2}
E= mc^2 = \rho V c^{2},
\ee
where $\rho$ is the mass density and $\varepsilon \equiv \rho c^2$ is the energy density, then we can take the differentials of $E$ getting
\citep[]{Youm2002}
\be
dE= c^{2} \rho dV +c^{2}V d \rho +2 \rho V c dc ,
\ee
which inserted into (\ref{1stlaw}) and taking the time derivative gives
\be
\label{1stt}
\dot{\rho} + \frac{\dot{V}}{V} \left( \rho + \frac{p}{c^2} \right) +  2 \rho \frac{\dot{c}}{c} - \frac{T}{V c^{2}} \dot{S} = 0~.
\ee
On the other hand, the continuity equation for varying $c$ models gives
\be
\label{Bianchi}
\dot{\rho} + \frac{\dot{V}}{V} \left( \rho + \frac{p}{c^2} \right) =\frac{3k c \dot{c}}{4 \pi G a^{2}}~.
\ee
Comparison of (\ref{1stt}) and (\ref{Bianchi}) leads to
\be
\label{equiv}
2 \rho \frac{\dot{c}}{c} - \frac{T}{V c^{2}} \dot{S} = - \frac{3k c \dot{c}}{4 \pi G a^{2}}~.
\ee
Defining
\bea
\label{wrho}
\widetilde{\rho} &=& \frac{3}{8 \pi G} \left( \frac{\dot{a}^2}{a^2} + 2 \frac{k c^2}{a^2} \right) = \rho + \frac{3}{8 \pi G} \frac{k c^2}{a^2}~,\\
\label{wp}
\widetilde{p} &=& p - \frac{1}{8 \pi G} \frac{k c^4}{a^2}~,
\eea
one can rewrite (\ref{equiv}) assuming that the pressure and the mass density are proportional
\be
\label{tpw}
\tilde{p} = \tilde{w} \tilde{\varrho} c^2
\ee
as
\be
\label{dotSV}
\dot{S} = 2 \frac{\widetilde{\rho} V c^{2}}{ T} \frac{\dot{c}}{c}~.
\ee
From (\ref{dotSV}) we note a very nice analogy to an ideal gas equation of state provided that we take
\be
\label{NkBc}
\frac{ \widetilde{\rho} V c^{2}}{ T} = {\rm const.} = \frac{N k_B}{\tilde{w}},
\ee
where $N$ is the number of particles, $k_B$ is the Boltzmann constant. After integrating (\ref{dotSV}), it is easy to define the entropy in a form which is analogous to the standard Gibbs relation (it is really interesting to note that in this case the volume of the phase space is proportional to the speed of light)
\be
\label{Sc}
S(t) = 2\frac{N k_B}{\tilde{w}} \ln{\left[ A _{0} c(t) \right]},
\ee
where we have included an integration constant $A_{0}$ which, without the loss of generality can be taken to be one.
All this simplifies for flat $k=0$ models since for them $\widetilde{\rho} = \rho$, and using the barotropic equation of state $p = w \rho c^2$ with the barotropic index $w=$ const. we have
\be
\label{Sc1}
S(t) = \frac{2}{w} \frac{pV}{T} \ln{c(t)} = \frac{2}{w} N k_B \ln{c(t)}~~.
\ee

\subsection{thermodynamics of varying $G$}

For the varying gravitational constant $G$ ($\dot{c} =0$) models, one starts with the continuity equation (\ref{conser}) (which is an equivalent of the first law) to get
\be
\dot{\rho} + \frac{\dot{V}}{V} \left( \rho + \frac{p}{c^2} \right) =- \rho \frac{\dot{G}}{G},
\ee
so that
\be
\dot{S}= -\frac{\rho V c^{2}}{T} \frac{\dot{G}}{G},
\ee
and again assuming an ideal gas equation of state
\be
\frac{ \rho V c^{2}}{T} = {\rm const.} = N k_B ,
\ee
one gets for varying $G$ that
\be
\label{SG}
 S(t) = N k_B \ln{\left[\frac{A_0}{G(t)} \right]},
 \ee
where $A_0$ is an integration constant (and can be taken to be one).

\subsection{varying $c$ cyclic doubleverse}
\label{5.3}

From (\ref{Sc}) and (\ref{SG}) one can see that the entropy is a function of the fundamental constants and it can increase or decrease depending on the form of the functions $c(t)$ and $G(t)$. According to the second law of thermodynamics entropy should never decrease and so we have to pick up some restricted form of these functions. However, this should be the case in our universe only. In the multiverse composed of individual universes this can be different. Our idea is that the total entropy of such an entity can still be constant or increase, while the entropies of individual universes can either increase or decrease.

\begin{figure}
%[htbp]
\includegraphics[width=8.3cm]{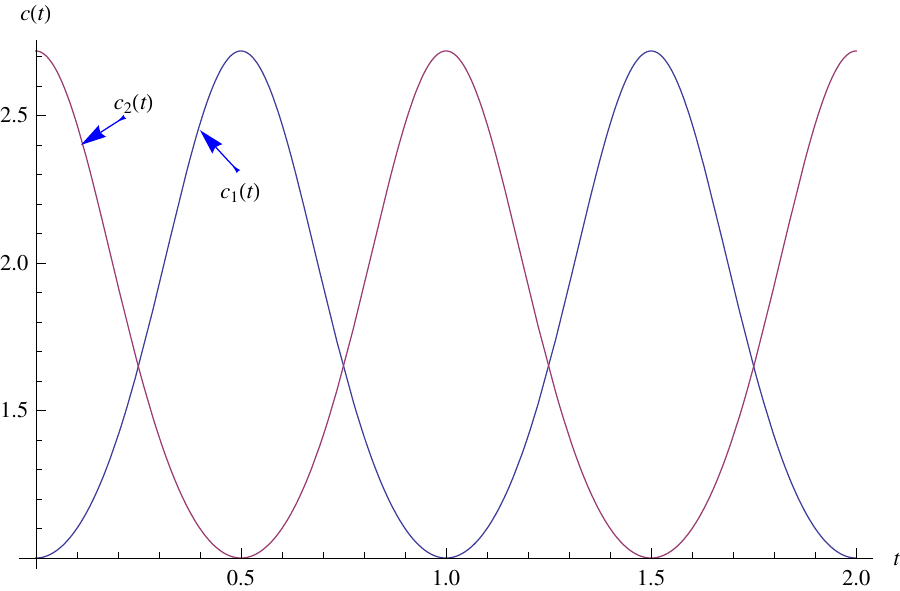}
\includegraphics[width=8.3cm]{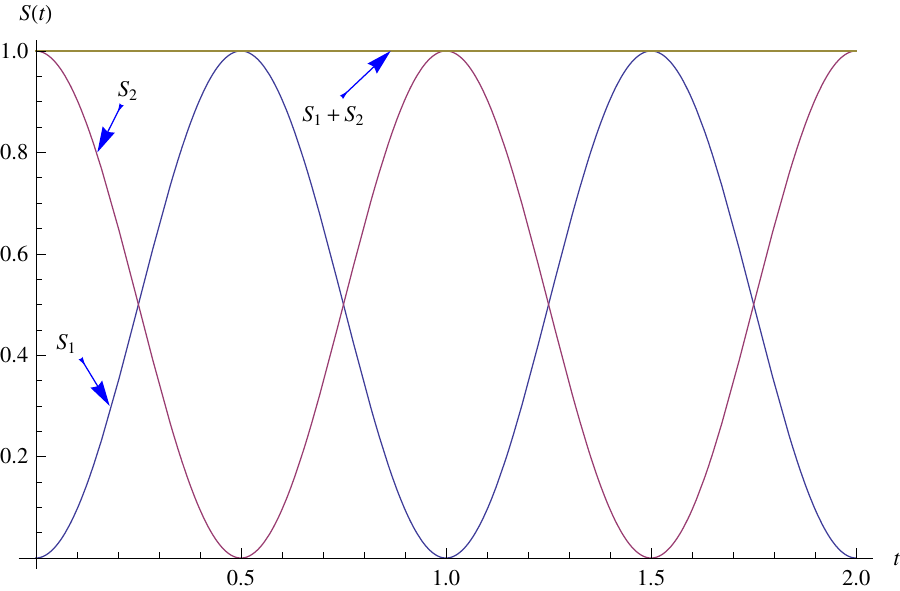}
\caption{The speeds of light (\ref{cfun1}),  (\ref{cfun2}) and the entropies (\ref{ec1})-(\ref{ec2}) of the varying $c$ doubleverse.}
\label{plotc}
\end{figure}

Let us assume that the entropy of the multiverse is constant. Since the entropy is an extensive quantity, then we have
\be
\label{sumS}
\dot{S} = \sum_{i=1}^{n} \dot{S}_{i} = \dot{S}_{1}+\dot{S}_{2}+\dot{S}_{3}+...+\dot{S}_{n} = 0 .
\ee
Specifically, for two universes which form a multiverse (which from now on we will call the ``doubleverse'')  from (\ref{Sc1}) we have:
\bea
\label{en1}
S_{1}= \frac{2}{\tilde{w}} N_1 k_B \ln {\left[ c_{1}(t) \right]}, \\
\label{en2}
S_{2}= \frac{2}{\tilde{w}} N_2 k_B \ln {\left[ c_{2}(t) \right]}.
\eea
We assume the following ans\"atze for the functions $c(t)$:
\bea
\label{cfun1}
c_{1} \left( t \right) = e^ {\lambda_{1} \phi_{1} \left( t \right)}, \\
\label{cfun2}
c_{2} \left( t \right) = e^ {\lambda_{2} \phi_{2} \left( t \right)},
\eea
where $\lambda_1$ and $\lambda_2$ are constants. The total entropy of the doubleverse is
\be
\label{ec12}
S=S_{1}+ S_{2} = 2  \frac{ \widetilde{\rho_{1}} V_{1} c_{1}^{2}}{ T_{1}} \lambda_{1} \phi_{1}
+ 2 \frac{ \widetilde{\rho_{2}} V_{2} c_{2}^{2}}{ T_{2}} \lambda_{2} \phi_{2} \left( t \right)~.
\ee
An interesting case which can be solved exactly is when
\be
\label{lamrel1}
\frac{\tilde{\varrho}_{1} V_{1} c_{1}^{2}}{ T_{1}} \lambda_{1}=  \frac{\tilde{\varrho}_{2} V_{2} c_{2}^{2}}{ T_{2}} \lambda_{2}~,
\ee
or on the footing of (\ref{NkBc})
\be
N_1 \lambda_1 = N_2 \lambda_2~,
\ee
which allows to pick up the following functional dependence of $c_1$ and $c_2$
\bea
\phi_{1} \left( t \right)=\sin^2{\left( \pi \frac{t}{t_s}\right)}, \\
\phi_{2} \left( t \right)=\cos^2{\left( \pi \frac{t}{t_s}\right)},
\eea
and write (\ref{en1}) and (\ref{en2}) as
\bea
\label{ec1}
S_1 &=& \frac{2}{\tilde{w}} N_1 k_B \lambda_1 \sin^2{\left( \pi \frac{t}{t_s}\right)}, \\
\label{ec2}
S_2 &=& \frac{2}{\tilde{w}} N_2 k_B \lambda_2 \cos^2{\left( \pi \frac{t}{t_s}\right)}.
\eea
The plots of (\ref{cfun1})-(\ref{cfun2}) and (\ref{ec1})-(\ref{ec2}) are given in Fig. \ref{plotc}.

An example of a doubleverse model which fulfils the field equations (\ref{wrho}), (\ref{wp}), and (\ref{tpw}) is the one with 
$\tilde{w}=- 1/3$, which corresponds to already mentioned cosmic strings universe for which the scale factors are
\be
\label{at}
a(t) = a_1(t) = a_2(t) = a_0 t .
\ee
On the other hand, assuming that still the pressure and the energy density are proportional
\be
\frac{\tilde{p}(t)}{c^2 \left( t \right)} \sim \tilde{\rho}(t) ,
\ee
we can extend (\ref{tpw}) with $\tilde{w} = -1/3$ into the form
\be
\label{eosc}
\frac{\tilde{p}(t)}{c^2  \left( t \right)} = - \frac{1}{3} \tilde{\rho}(t) +\frac{p_0(t)}{c^2 \left( t \right)} ,
\ee
where
\be
\label{p0}
\frac{p_0(t)}{c^2(t)} = -\frac{1}{4\pi G} \frac{\ddot{a}}{a} ,
\ee
and so (\ref{NkBc}) transforms into
\be
\frac{ \left(\widetilde{p} - p_0 \right) V}{T} = const.
\ee
which is an equation of state in the form of the van der Waals imperfect gas model. 

Now, one may introduce the two scale factors equal and of the similar form as in (\ref{acyclic1}):
\be
\label{a1a2}
a(t) =  a_1 \left( t \right) =  a_2 \left( t \right) = a_0 \left| \sin{\left( \pi \frac{t}{t_s}\right)} \right|
\ee
which satisfy (\ref{wrho}), (\ref{wp}) and the  equation of state (\ref{eosc}) with a constant $p_0/c^2 = \pi/(4Gt_s^2)$. Then, the evolution is evidently cyclic. The scale factor (\ref{at}) is recovered from (\ref{p0}) provided that $p_0=0$. The point here is that although the geometrical evolution of the ''parallel'' universes is the same, this is not the case with {\it the evolution of the physical constants $c_1(t)$ and $c_2(t)$ which is evidently different}. 

\subsection{varying $G$ cyclic doubleverse}

For varying $G$ models we use (\ref{SG}) to get:
\bea
\label{s1s2}
S = S_1 + S_2 = \frac{\rho_{1} V_{1} c^{2}}{T_{1}} \ln{\left(\frac{A_1}{G_{1} \left( t \right)} \right)}+ \frac{\rho_{2} V_{2} c^{2}}{T_{2}} \ln{\left(\frac{A_2}{G_{2} \left( t \right)} \right)} .\nonumber \\
\eea
We choose the relation between the gravitational coupling constant $G(t)$ and the scale factor $a(t)$ (taken in the form given by (\ref{Gcyclic1})) in the first universe as
\be
\label{g1}
G_{1}(t) = G_{A}/a_1^2(t) ,
\ee
where $a_1(t)$ is the scale factor of the first universe, while in the second universe we choose that
\be
\label{g2}
G_{2}(t) = G_{B} a_2^2(t) ,
\ee
where $a_2(t)$ is the scale factor of the second universe. In order to compare different evolutions of the varying physical constants $G_1$ and $G_2$, as the first attempt, we assume that both scale factors are equal i.e.
\be
a_1(t) = a_2(t) \equiv a(t).
\ee
In other words, both universes have the same geometrical evolution. In order to get the total entropy constant, we should again apply the condition similar to (\ref{lamrel1}):
\be
\frac{\rho_{1} V_{1} c^{2}}{T_{1}} \widetilde{A_{1}} = \frac{\rho_{2} V_{2} c^{2}}{T_{2}} \widetilde{A_{2}}~,
\ee
where
\be
\widetilde{A_{1}}= \ln{ \left( \frac{ A_1}{ G_{A}} \right)}, \hspace{0.5cm} \widetilde{A_{2}}= \ln{A_2 G_{B}} .
\ee
Then, we have two universes with the total (net) entropy constant, though it is decreasing or increasing in each individual universe. Comparing the scale factors in both universes we have that
\be
G_1(t) \propto \frac{1}{G_2(t)} ,
\ee
and
\be
\label{S1S2}
S_1(t) \propto \ln{G_1(t)} \propto \ln{\frac{1}{G_2(t)}} \propto \frac{1}{S_2(t)} .
\ee

\begin{figure}
%[htbp]
\includegraphics[width=8.3cm]{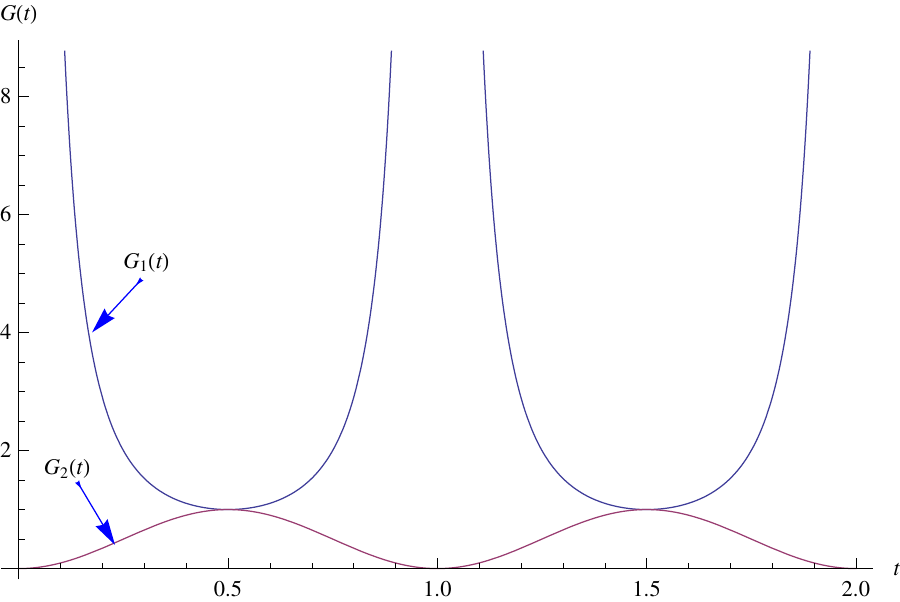}
\includegraphics[width=8.3cm]{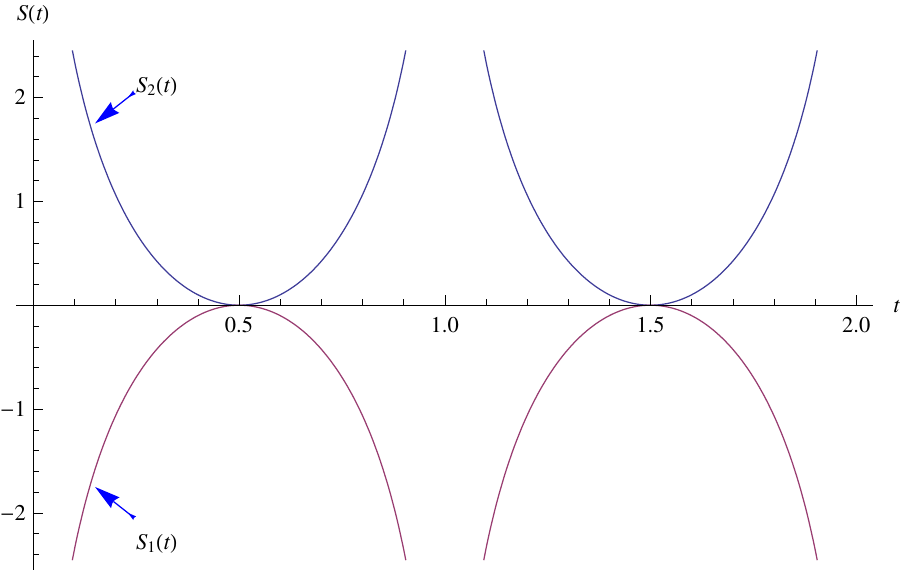}
\caption{Gravitational constants (\ref{g1}),  (\ref{g2}) and the entropies (\ref{s1s2}) of the doubleverse for the scale factor (\ref{a1a2}).}
\label{G1G2S1S2}
\end{figure}

This means that we have the same geometrical evolution of the two ``parallel'' universes, but their physical evolution (physical coupling constants) is different. More precisely, the first universe and the second universe {\it have the same scale factors but their gravitational constants are different}. In one of these universes the evolution of the gravitational constant is singular depending on the evolution of the scale factor $a(t)$. This is an alternative approach than in the case of the Kaluza-Klein theory in which microscopic dimensions shrink and have the scale factor which diminishes while macroscopic dimensions expand \citep[]{Polchinski1998}.

\subsection{discussion}

For the whole body of this Section it is also worth mentioning that the scale factors for each universe within the multiverse satisfy the field equations (\ref{rho}) and (\ref{p}). 

From Fig.\ref{G1G2S1S2}, which has been plotted under the assumption that $G_A = G_B$ and $a_0 = 1$, one can see that the universe 1 may replace its evolution along the trajectory of the universe 2 {\it at the maximum expansion point}. It is an interesting new option for the evolution of the universe -- now put in the context of the multiverse -- where some effects take place at the turning point. Similar effects (though considered to be appropriate to {\it the same} universe) were studied in the context of quantum cosmology and the probability of tunneling from an expanding phase into a harmonic oscillator phase was calculated \citep[]{Dafmmodeboxclsecibrowski1995}. This discussion was more recently developed under the name of the simple harmonic universe (SHU) scenario \citep[]{Mithani2012,Mithani2014}. All this is in agreement with the claim that macroscopic quantum effects in cosmology are possible \citep[]{Gold1962,Kiefer1988,Kiefer1995,Dabrowski2006}.

The same is possible for the varying $c$ multiverse of subsection \ref{5.3} where at the maximum expansion of the scale factors (\ref{a1a2}) the universes may replace their evolution due to quantum effects. 

Here we mean the multiverse which is quantum mechanically entangled and there are periods of its evolution where this entanglement matters (e.g. the maximum expansion point) turning and enters the behaviour of individual universes, but most of the evolution of the individual universes is classical.  In other words, we try to say something about possible parallel evolutions of the two classical patches of the multiverse (which can be understood on some lower levels of the hierarchy given by \citet{Tegmark2003} - from I to IV). These two patches are classically disconnected, but they can be quantum mechanically entangled \citep[]{Robles-Perez2010,Robles-Perez2014,Robles-Perez2015} and the effect of entanglement can be imprinted in one of the universes (for example in CMB) or become explicit when these universes reach some special conditions (as we have proposed for example - at the maximum expansion).

\section{Cyclic Brans-Dicke multiverse}
\setcounter{equation}{0}
\label{cyclicBD}

Brans-Dicke field equations \citep[]{Brans1961} can be written down as (and following Section \ref{CyclicMulti} are valid in each individual universe of the multiverse separately giving the same form of the field equations (\ref{rho}) and (\ref{p}) in the limit $\phi=$ const.)
\bea
\label{BDrho}
\rho \left( t \right) &=& \frac{3 \phi \left( t \right) \left[ \dot{a}^2  \left( t \right)+k c^2 \right]} {8 \pi a^2  \left( t \right)} + \frac{3 \dot{a} \left( t \right) \dot{\phi} \left( t \right)} {8 \pi a \left( t \right)}-\frac{\omega \dot{\phi}^2 \left( t \right)}{16 \pi \phi  \left( t \right)}, \\
\label{BDp}
p \left( t \right) &=& - \frac{c^2}{8 \pi} \left[\frac{\phi \left( t \right) \left[ \dot{a}^2  \left( t \right) +k c^2 \right]}{ a^2  \left( t \right)} - \frac{ 2 \dot{a} \left( t \right) \dot{\phi} \left( t \right)} { a \left( t \right)}-\frac{\omega \dot{\phi}^2 \left( t \right)}{2 \pi \phi  \left( t \right)} \right. \nonumber \\
&-& \left. \frac{2 \phi \left( t \right) \ddot{a} \left( t \right)}{ a  \left( t \right)}-\ddot{\phi} \left( t \right) \right] ,\\
\label{BDpsi}
\Box\phi &=& \frac{8 \pi}{3+2 \omega} T ,
\eea
and the last of them (\ref{BDpsi}), absent in the system (\ref{rho})-(\ref{conser}), is equivalent to the continuity equation in Einstein relativity
\be
\label{BDcons}
\dot{ \rho } + \frac{\dot{V}}{V} \left( \rho + {\frac{p}{c^2}} \right) =0 ,
\ee
where one can take $V = a^{3}$, which describes a model of constant entropy. It also differs from the conservation equation of varying $G$ theory (\ref{conser}) since it does not contain the term $-\varrho \dot{G}/G$ which couples mass density with variation of gravitational interaction. 

However, if we want to assume that the entropy is not constant, then we need an extra entropic term in the continuity equation, getting
\be
\label{rhoS}
\dot{ \rho } + \frac{\dot{V}}{V} \left( \rho + {\frac{p}{c^2}} \right) - 16 \pi \frac{T \dot{S}}{V} \frac{\phi}{\dot{\phi}} =0 .
\ee
This leads to a modification of the field equation (\ref{BDpsi}) in Brans-Dicke theory by an additional function (we take $k=+1$ in (\ref{BDrho})-(\ref{BDp}))
\be
\label{BoxN}
\Box \phi=\frac{8 \pi}{3+2 \omega} T + N \left( t \right) ,
\ee
and the function $N \left( t \right) $ is responsible for an increase or a decrease of the entropy.

Such modification has recently been considered by \citet[]{Kofinas2015}  in the context of energy exchange between a scalar field and the standard matter. This allows an avoidance of having the problem with singular conformal cosmology (and also cyclic conformal cosmology - CCC \citep[]{Penrose2009book,Penrose2010,Majid2008}) limit $\omega = -3/2$  (or Kofinas' $\lambda = 2/(3+2\omega) \to \infty$) and allows the entropy growth needed to drive our cyclic multiverse models \citep[]{Blaschke2012}.

Comparing (\ref{rhoS}) and (\ref{BoxN}), one gets
\be
\label{dotS}
\dot{S}= \frac{N \left( t \right) V}{16 \pi T} \frac{\dot{\phi}}{ \phi} .
\ee
Since the mass density and the pressure are both the functions of time, then we can make $N \left( t \right)$ the function the of the mass density and the pressure $N(p,\rho)$ and assume that
\be
\frac{ N \left( p , \rho \right) V}{T} = const.
\ee
For variable entropy, one modifies the continuity equation (\ref{BDcons}) to the following form
\be
\ddot{\phi} + \frac{3 \dot{a} \left( t \right) \dot{\phi}  \left( t \right)}{a \left( t \right)} =
\frac{8 \pi \left[\rho  \left( t \right) - 3 p  \left( t \right) \right]}{3+2 \omega} + N(t) ,
\ee
where
\be
N(t) = \frac{8 \pi \alpha \rho \left( t \right) }{3+2 \omega} +  \frac{8 \pi \beta p \left( t \right) }{3+2 \omega},
\ee
and $\alpha$ and $\beta$ are some constants.

In such a case the entropy equation (\ref{dotS}) integrates to give
\be
%\dot{S} &=& \frac{ \left[ \alpha \rho \left( t \right)  +  \beta p \left( t \right) \right] V}{2T \left( 3+2 \omega \right) } \frac{\dot{ \phi}}{\phi} ,\\
S =  \frac{ \left[ \alpha \rho \left( t \right)  +  \beta p \left( t \right) \right] V}{2T  \left( 3+2 \omega \right) } \ln{\phi},
\ee
where for $ \alpha \neq 0$ we need to use an equation of state $p \sim \rho $.

In a special case of the low-energy-effective-string theory \citep[]{Gasperini2003,Lidsey2000}, where $\omega = -1$ and $ k=+1 $, one can take the ansatz (\ref{acyclic1}) in the first universe (of the doubleverse) as 
\be
\label{a1BD}
a_1 \left( t \right) = a_A \left| \sin{\left( \pi \frac{t}{t_s} \right)} \right| ,
\ee
and
\be
\label{phi1BD}
\phi_1 \left( t \right) \sim  a_1^2 \left( t \right) = \phi_A \sin^2{\left( \pi \frac{t}{t_s} \right)} ,
\ee
where $ \alpha_1 = 4/5$, $ \beta_1 = - 32/15$ and the speed of light $ c = (a_A \pi)/(t_s)$  necessarily in order to fulfill the field equations (\ref{BDrho})-(\ref{BDpsi}). In this case the mass density and the pressure read as
\bea
\rho_1(t) &=&  \frac{ \phi_A \pi}{8 {t_s}^2} \left[ 3 + 11\cos^2{\left( \pi \frac{t}{t_s} \right)} \right] ,\\
p_1(t) &=&  \frac{ c^2 \phi_A \pi}{8 {t_s}^2} \left[ 3 - 9 \cos^2{\left( \pi \frac{t}{t_s} \right)} \right],
\eea
while the equation of state is
\be
\frac{p_1 \left( t \right)}{c^2} = - \frac{9}{11} \rho_1 \left( t \right) + \frac{ 15 \phi_A \pi}{ 22 {t_s}^2}.
\ee
In the second universe (out of the doubleverse), we can assume that
\be
\label{a2BD}
a_2 \left( t \right) = a_B \left| \sin{\left( \pi \frac{t}{t_s} \right)} \right|,
\ee
and
\be
\label{phi2BD}
\phi_2 \left( t \right) \sim  \frac{1}{a_2^2 \left( t \right)} = \frac{\phi_B}{ \sin^2{\left( \pi \frac{t}{t_s} \right)}} ,
\ee
with $ \alpha_2 = 0 $ and $ \beta_2 =-4 $ and the speed of light $ c = (a_B \pi)/t_s$.
In this model, the mass density and the pressure take the form
\bea
\rho_2 \left( t \right) &=& \frac{ \phi_2 \pi}{8 {t_s}^2} \left[ \frac{2+ \sin^2\left( \pi \frac{t}{t_s} \right)}{\sin^4\left( \pi \frac{t}{t_s} \right)} \right] , \\
p_2 \left( t \right) &=& - \frac{c^2 \phi_2 \pi}{8 {t_s}^2} \left[ \frac{2- \sin^2\left( \pi \frac{t}{t_s} \right)}{\sin^4\left( \pi \frac{t}{t_s} \right)} \right] ,
\eea
and we used the fact that for $ \alpha_2 = 0$ the equation of state is not necessary.
Finally, provided that
\be
\ln{\phi_A} \frac{ \left[ \alpha_1 \rho_1 \left( t \right)  +  \beta_1 p_1 \left( t \right) \right] V_1}{T_1  } = \ln{\phi_B}
\frac{ \left[ \alpha_2 \rho_2 \left( t \right)  +  \beta_2 p_2 \left( t \right) \right] V_2}{T_2  } ,
\ee
the sum of entropies of individual universes in such a doubleverse is constant i.e. $\dot{S} = 0$ (cf. Eq. \ref{sumS}), though it changes in each of them. 

As we can see from (\ref{a1BD}), (\ref{a2BD}) and (\ref{phi1BD}, (\ref{phi2BD}) the geometrical evolution of the scale factors is similar and cyclic as in the previous models of Sections \ref{cyclic} and \ref{CyclicMulti}, while the evolution of the physical constants represented here by the dynamical fields $\phi_1$ and $\phi_2$ which play the role of $G_1$ and $G_2$ can be different. In other words, all the discussion of the physical effects of Section \ref{CyclicMulti} is valid. 

\section{Conclusions}
\setcounter{equation}{0}
\label{conclusion}

In this paper we have extended our idea of regularisation of cosmological singularities due to the variability of the fundamental constants \citep[]{Dabrowski2013} onto the cyclic universe and multiverse models. We have shown new examples of regularisation of the sudden future singularity by varying $c$ and $G$ though continuing towards a little-rip singularity which prevented the construction of cyclic universes.

Then, by an appropriate choice of the evolution of the scale factor $a(t)$ and the dynamics of $G(t)$, we have been able to construct cyclic models in which the energy density and the pressure are non-singular and oscillatory.

Being encouraged by the success of such a scenario we have extended this idea onto the multiverse containing cyclic individual universes with either growing or decreasing entropy, though leaving the net entropy constant. In particular, we have considered the ``doubleverse'' with the same geometrical evolution of the two ``parallel'' universes with the evolution of their physical coupling constants $c(t)$ and $G(t)$ being different. We have found an interesting possibility for one universe (of the doubleverse) to exchange its evolution history with another universe at the turning point of expansion. Similar investigations have been made in the context of quantum cosmology, where the universe (considered as the same object/entity) could quantum mechanically tunnel into an oscillatory (harmonic) regime.

We have also found that similar scenario is possible within the framework of Brans-Dicke theory of varying $G(t) \propto 1/\phi(t)$, where $\phi(t)$ is the Brans-Dicke field though these two theories differ in the form of the continuity equation. 

In fact, we face various categories of the ''bounces''. We have differentiated the "singular bounces'' and ''non-singular-bounces'' referring to the appropriate quantities such as the scale factor, mass density and pressure. We have also calculated the energy conditions and noticed that the null energy condition is not necessarily violated due to the regularisation by variability of $G$. The discussion of the energy conditions shows that the null energy condition is fulfilled for the big-bang type bounces and violated for the big-rip type bounces. 

Our discussion is heuristic and preliminary so we are not able to address all the physical problems which are related to the issue of cyclic models within this scenario. Besides, we stick to purely isotropic geometry models not discussing the problem of possible anisotropy domination which can be solved by the very stiff equation of state ($p = w \rho$, $w \gg 1$) during the collapse phase. We have explicitly found analogies with ekpyrotic/cyclic approach where the role of varying $G$ is played by the appropriate coupling of matter with a scalar field. The  ghost (phantom) instabilities may not be faced here (at least partially) mainly because in most of the cases still there is no null energy condition violation. 

Finally, it should be said that the main problem in the literature related to the multiverse idea concentrates on the measure problem, but we have not  investigated it in any aspect here (see e.g. \citet[]{Lehners2012}). Instead, we try to say something about possible parallel evolutions of the two classical patches of the multiverse (which can be understood on some lower levels of the hierarchy I to IV of \citet[]{Tegmark2003}). Individual universes/patches of the universe are classically disconnected, but they can be quantum mechanically entangled and the effect of entanglement can in principle be imprinted in individual universes or become explicit when these universes reach some special conditions. 

\section{Acknowledgements}

This work was supported by the National Science Center Grant DEC-2012/06/A/ST2/00395. We thank anonymous referee for constructive suggestions.

\bibliographystyle{mnras}
\bibliography{citations5}  % if your bibtex file is called example.bib

%\end{document}
%\begin{thebibliography}{99}
%\end{thebibliography}

\end{document}